\documentclass[12pt,a4paper]{article}
\usepackage[utf8x]{inputenc}
\usepackage{ucs}

\usepackage{graphicx}
\usepackage[left=1.5cm,right=1.5cm,top=2cm,bottom=2cm]{geometry}

\usepackage[bookmarks=true,colorlinks=true,citecolor=blue,linkcolor=red,urlcolor=magenta]{hyperref}

\usepackage{amsmath} 
\usepackage{amssymb} 

\usepackage{multirow}
\usepackage[table]{xcolor}

\usepackage{multicol}

\usepackage{float}

\usepackage[numbers,sort&compress]{natbib}

\title{FAST TRAVELING-WAVE REACTOR OF THE CHANNEL TYPE}

\author{Vitaliy D. Rusov$^1$\footnote{Corresponding author E-mail: siiis@te.net.ua}, Victor A. Tarasov$^1$, Volodymyr N. Vashchenko$^2$,\\
             Sergei A. Chernezhenko$^1$, Andrei A. Kakaev$^1$, Oksana I. Pantak$^1$}

\date{}

\begin{document}

\maketitle

\begin{center}
$^1$Department of Theoretical and Experimental Nuclear Physics, \\
Odessa  National Polytechnic University, Odessa, Ukraine

$^2$State Ecological Academy for Postgraduate Education, Kiev, Ukraine
\end{center}

\begin{abstract}
The main aim of this paper is to solve the technological problems of the TWR based on the technical concept described in our priority of invention reference~\citep{ref01},  which makes it impossible, in particular, for the fuel claddings damaging doses of fast neutrons to excess the ~200 dpa limit. Thus the essence of the technical concept is to provide a given neutron flux at the fuel claddings by setting the appropriate speed of the fuel motion relative to the nuclear burning wave.

The basic design of the fast uranium-plutonium nuclear traveling-wave reactor with a softened neutron spectrum is developed, which solves the problem of the radiation resistance of the fuel claddings material.
\end{abstract}

\section{Introduction}

Today the idea of a wave-like neutron-nuclear burning is almost undisputed. Meanwhile in the field of physical theory, in our opinion, there are some problems still unexplored and extremely important. Among them are the influence of the heat transfer during the temperature and pressure change over a wide range, the phase state of the fissile medium and its influence on the existence and stability of a nuclear burning wave. Such problems as the heterogeneous structure of the core, the influence of the radiation-induced defects kinetics in the fuel, the heat convection and mixing (liquid or gas fuel), the radiation resistance of the fuel claddings construction materials, the ignition modes (initialization) and others also remain unexplored. It is also interesting to study the kinetics of the neutron-nuclear burning in combined fissile media  (uranium-plutonium, uranium-thorium medium with various pre-enrichments in $^{258}$U, $^{233}$U, $^{239}$Pu and possibly some other fissile nuclides such as $^{241}$Pu or Cm) and, consequently, the combined uranium-plutonium and thorium-uranium burning waves, and perhaps even some others, as well as the kinetics of the nuclear burning wave  reflection from the boundaries of the medium (neutron reflector), the repeated waves and the  burning waves interference.

At the same time, some of the technological problems of TWR are very actively discussed in the scientific community today. This often leads to a conclusion about the impossibility of such project~\citep{ref02} because of a number of its insurmountable disadvantages:

\begin{itemize}
	\item
	a high degree of nuclear fuel burn-out (at least 20\% on average), which assumes:
	\begin{itemize}
		\item a high damaging dose from fast neutrons acting on the material of the fuel claddings (~500 dpa);
		\item large gas emission, which requires increasing of the inner gas cavity length in the  long fuel rods;
	\end{itemize}
	\item a long active zone, which requires the use of the long fuel rods, the parameters of which are unacceptable from the exploitation point of view. In particular, this applies to the parameters characterized by significant growth of:
	\begin{itemize}
		\item the value of the positive void coefficient of reactivity;
		\item the hydraulic resistance;
		\item the energy consumption for pumping of the coolant through the reactor;
	\end{itemize}
	\item The problem of spent nuclear fuel, associated with the need of the unburned plutonium processing and disposal of the radioactive waste.
\end{itemize}

However nowadays there are several proposals of the possible principal TWR designs~\citep{ref03,ref04a,ref05,ref06,ref07,ref08,ref09,ref10,ref11,ref12,ref13,ref14} based on scientific results~\citep{ref15,ref16,ref17,ref17a,ref18,ref19,ref20,ref21,ref22,ref23,ref24,ref25,ref26,ref27}, proving the possibility of the theoretical and technical implementation of the slow neutron-nuclear burning modes.

A wave reactor on fast neutrons in self-regulating neutron-nuclear mode of the second kind is known. First it was presented in papers~\citep{ref05,ref06} and its advanced implementation was patented in~\citep{ref14}.

Let us consider its drawbacks.

This reactor does not implement a slow traveling soliton-like wave of neutron-nuclear burning mode. All the volume of the reactor active zone during its operation is in self-regulating neutron-nuclear mode of the second kind. Such reactor is not completely self-regulating, but requires some criticality regulation by the control systems. The reactor active zone consists of two or several zones, one of them being a neutron source, providing the necessary neutron production at its start and therefore contains enriched reactor fuel requiring supercritical load and absorbing control rods and a protection system. The possible $^{238}$U burn-out in non-enriched zones ("breeding" fuel zone) does not exceed 10\%.

A wave reactor on fast neutrons is also known~\citep{ref08,ref09,ref10,ref11,ref12,ref13}. It was proposed by "Terra-power"  to operate in the so-called standing wave mode.

Its drawbacks:

The slow traveling soliton-like neutron-nuclear burning is not implemented in this reactor. The entire volume of the reactor active zone represents a neutron-nuclear burning zone during its operation. The reactor active zone consists of two zones, the central zone being a neutron source and provides the necessary neutron generation during reactor start-up and operation, and therefore contains enriched reactor fuel, creating supercritical load and contains absorbing control and protection system rods for reactor control. During the reactor start-up and some time after the start-up the reactor criticality regulation by external control systems is required. Such adjustment is analogous to operation modes of the usual fast reactors with small excessive reactivity and requires the similar control and protection systems. Therefore, in reactors proposed by "Terra-power" like in reactors proposed in~\citep{ref14}, a significant advantage of traveling burning wave reactor is absent – a complete self-regulation of the reactor active zone and consequently, the possibility of significant simplification and cheapening of the control and protection systems. A solution of the excessive shell materials damaging dose problem is also absent in reactors proposed by "Terra-power".

It should be noted however, that the "Burns and Roe" company, specializing in the design and construction of the nuclear power stations, already offers the architectural and engineering projects for the conceptual design of TWR developed by "Terra-power" on their website.

In the papers by H.~Sekimoto~\citep{ref28,ref29,ref30,ref31,ref32,ref33,ref34,ref35,ref36,ref37,ref38,ref39,ref40,ref41,ref42,ref43,ref44,ref45,ref46,ref47,ref48,ref49,ref50,ref51,ref52,ref53,ref54,ref55,ref56,ref57,ref58,ref59,ref60,ref61,ref62,ref63,ref64}  a design of fast nuclear reactors capable of the traveling wave nuclear burning is studied. However, the main problem hindering the implementation of a traveling wave reactor -- the problem of the radiation damaging dose for the fuel claddings -- is also not solved in those studies.

H.~Sekimoto suggests an interesting idea~\citep{ref38,ref39,ref40,ref45,ref46} that the wave burning can be implemented in a thermal high-temperature gas reactor, for example, in the Japanese Experimental working reactor HTTR, if its fuel (enriched in $^{235}$U) previously added burnable neutron absorber $^{157}$Gd, it is interesting. And as presented in~\cite{ref48}, the results of mathematical modeling, wave fuel burning associated with its local transition in the field of burning in supercritical state, will be provided by local burnup absorber $^{157}$Gd in the burning zone (and in the initiation of the burning wave burnup absorber $^{157}$Gd in the nearest zone of nuclear fuel to an external source of neutrons).

\section{Fast TWR of the channel type with a fixed construction part}

In this section we present and justify a possible design of the fast TWR of channel type. As noted above, perhaps the most important is to solve the problem of high integral damaging dose of fast neutrons on construction materials in wave nuclear reactors, which may reach ~500dpa for the nuclear burning wave with the maximum burn-out of $^{235}$U. However, the materials capable of withstanding such a radiation load have not been created so far, and the maximum achievable radiation exposure for the reactor metals is 100 ÷ 200 dpa. Indeed, as shown in~\cite{ref02} and in Table~1 in~\citep{ref02}, none of the simulated burning wave modes  provide the necessary radiation resistance of construction materials in fuel claddings.

\begin{table*}
\rowcolors{3}{lightgray}{white}
\begin{center}
\begin{tabular}{|l|c|c|c|c|c|c|c|c|}
  \hline
  & $\Delta_{1/2}$ & $u$         & $\varphi$ & $\psi$ & $\left \langle \sigma_{dpa} \right \rangle$ & \multirow{2}{*}{$\dfrac{n_{dpa}}{200}$} & Fuel & \multirow{2}{*}{Solution} \\
  & [cm]                  & [cm/day] & [cm$^{-2}$s$^{-1}$] &  [cm$^{-2}$] & [barn] & &burn-up & \\
  \hline
  \multicolumn{9}{|c|}{\textbf{U-Pu cycle}} \\
  \hline
  Sekimoto 
  \cite{ref14a}  & 90 & 0.008 & 3.25 $\cdot$ 10$^{15}$ & 3.2 $\cdot$ 10 $^{23}$ & 1000 & 3.2 & $\sim$43\% & No \\
  \hline
  Rusov 
  \cite{ref29a} & 200 & 2.77 & 10$^{18}$ & 6.2$\cdot$10$^{24}$ & 1000 & 62 & $\sim$60\% & No\\
  \hline
  Pavlovich 
  \cite{ref26a} & -- & 0.003 & -- & 1.7$\cdot$10$^{24}$ & 1000 & 17 & $\sim$30\% & No\\
  \hline
  Fomin 
  \cite{ref15a} & 100 & 0.07 & 2 $\cdot$ 10$^{16}$ & 2.5$\cdot$10$^{24}$ & 1000 & 25 & $\sim$30\% & No\\
  \hline
  Fomin 
  \cite{ref13a} & 125 & 1.7 & 5 $\cdot$ 10$^{17}$ & 3.2$\cdot$10$^{24}$ & 1000 & 32 & $\sim$40\% & No\\
  \hline
  Chen 
  \cite{ref18a} & 216 & 0.046 & 3 $\cdot$ 10$^{15}$ & 1.2$\cdot$10$^{24}$ & 1000 & 12 & $\sim$30\% & No\\
  \hline
  Terra~Power 
  \cite{ref28a} & -- & -- & -- & -- & -- & 1.75 & $\sim$20\% & No\\
  \hline
  \multicolumn{9}{|c|}{\textbf{Th-U cycle}} \\
  \hline
  Teller 
  \cite{ref04a} & 70 & 0.14 & $\sim$2 $\cdot$ 10$^{15}$ & 8.6$\cdot$10$^{22}$ & 1000 & 0.96 & $\sim$50\% &  {\color{red} Yes}\\
  \hline
  Seifritz 
  \cite{ref5a} & 100 & 0.096 & 10$^{15}$ & 9.0$\cdot$10$^{22}$ & 1000 & 0.90 & $\sim$30\% &  {\color{red} Yes}\\
  \hline
  Melnik 
  \cite{ref25a} & 100 & 0.0055 & 0.5$\cdot$10$^{16}$ & 7.9$\cdot$10$^{24}$ & 1000 & $\sim$80 & $\sim$50\% & No \\
  \hline
  \multicolumn{9}{|c|}{\textbf{U-Pu (+ moderator)}} \\
  \hline
  Example & 100 & 0.234 & 2.5 $\cdot$ 10$^{15}$ & 9.2$\cdot$10$^{23}$ & 100 & 0.92 & $\sim$20\% &  {\color{red} Yes}\\
  \hline
  Ideal TWR & -- & -- & -- & 10$^{24}$ & 100 & 1.0 & $\geqslant$20\% &  {\color{red} Yes}\\
  \hline
\end{tabular}
\end{center}
\caption{Results of the numerical experiments of the wave mode parameters based 
         on U-Pu and Th-U cycles}
\label{tab1}
\end{table*}

The problem of the principal TWR design creation including the solution of the high integral damaging dose of fast neutrons in construction materials (first of all, the fuel rods shells) was set as a base for the invention. The specified problem is solved in this framework by a technical implementation of moving nuclear fuel (in which the nuclear burning wave travels) relative to the fuel rod shell. The movement speed is adjusted to provide the required integral fluence reduction at the fuel rod shell.

Let us consider the following ratio of the integral fluences for metals of the modern operating reactors and the TWR under development:

\begin{equation}
\dfrac{Fluence _{TWR} ^{metal}}{Fluence _{oper} ^{metal}} \sim \dfrac{\Phi_{TWR} \cdot t _{TWR}^{metal. camp.}}{\Phi_{oper} \cdot t _{oper}^{metal. camp.}} \sim \dfrac{500~dpa}{100~dpa}
\label{eq01}
\end{equation}

\noindent where $\Phi _{TWR}$ and $\Phi _{oper}$ are the neutron flux densities for the TWR and the modern operating reactors respectively. And 500~\textit{dpa} and 100~\textit{dpa} are the commonly considered values of the radiative resistance of the construction materials for possible nuclear burning modes with maximal $^{238}$U burn-out in TWRs and operating reactors respectively.

Let:
\begin{align}
\Phi _{oper} & \sim 10^{14} ~ neutrons / cm^2 \cdot s ~~ and ~~ t_{oper}^{camp} \sim 3 ~years \nonumber \\
\Phi _{TWR} & \sim 10^{17}~ neutrons / cm^2 \cdot s
\label{eq02}
\end{align}

Then from (\ref{eq01}) in case (\ref{eq02}) we obtain the estimate of the metals campaign time in TWR:

\begin{equation}
t _{TWR} ^{camp} \sim \frac{500 \cdot 10^{14} \cdot 3}{100 \cdot 10^{17}} years ~ \sim 1.5 \cdot 10^{-2} ~years ~ 
\sim 4.5 \cdot 10^5 ~s
\label{eq03}
\end{equation}

The time of the fuel nuclear burning in the active zone is

\begin{equation}
t_{fuel burn} = \frac{l_{fuel}}{v _{nucl. burn.}}
\label{eq04}
\end{equation}

\noindent where $l_{fuel}$ is the length of the active zone containing the fuel, $v_{nucl.burn.}$ is the nuclear burning wave traveling speed.

\begin{figure*}[htb]
\begin{center}
\includegraphics[width=12cm]{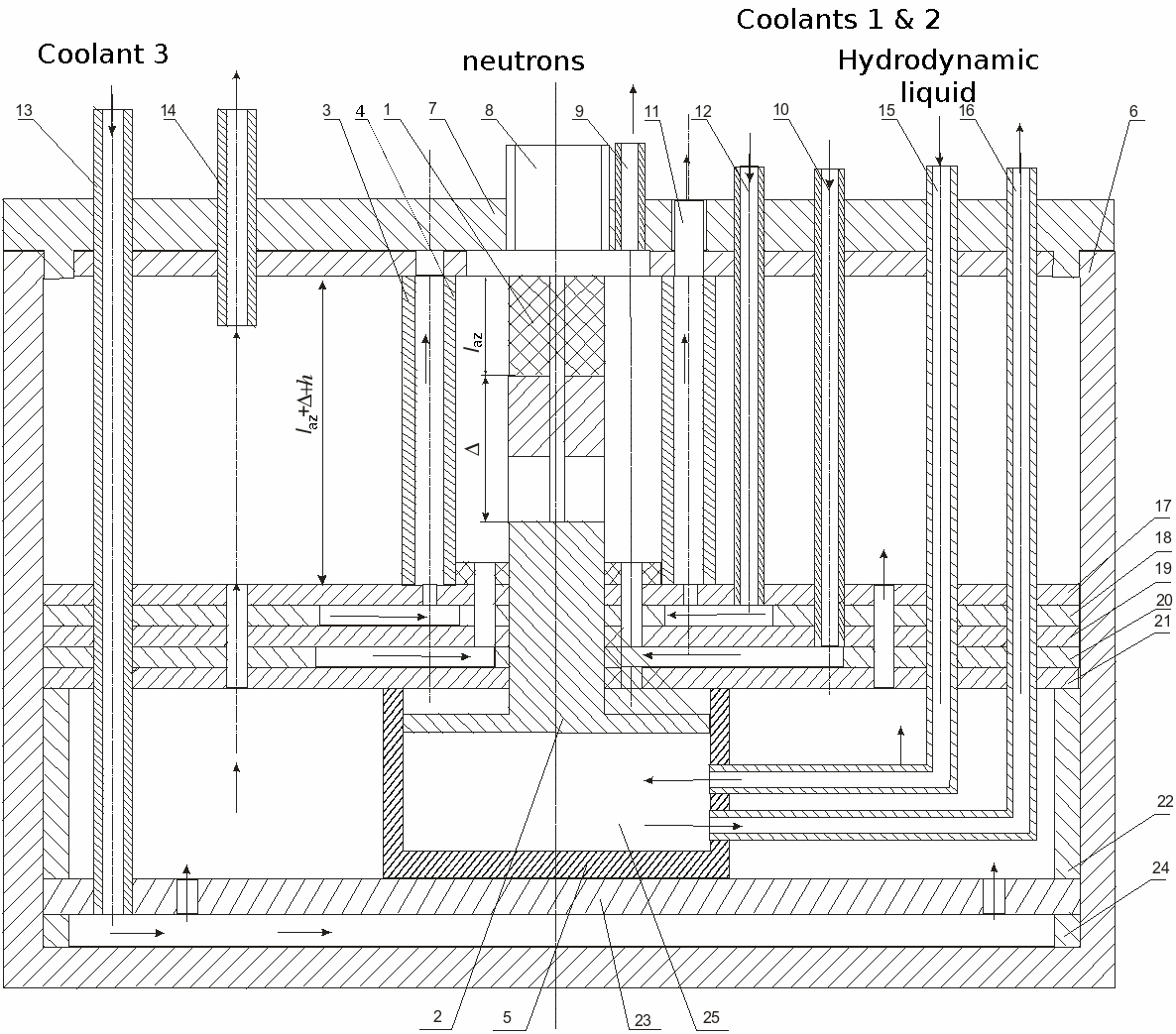}
\end{center}
\caption{Concept scheme of TWR prototype with one burning fuel rod. A variant of homogeneous active zone of large diameter ($\sim$1-3m).
(1 – the nuclear fuel; 2 – moving construction of the fuel movement hydraulic system; 3 – fuel channel shell; 4 - fuel rod shell-fuel channel shell; 5 – hydraulic liquid reservoir hull; 6 – reactor hull; 7 – reactor hull cover; 8 – neutron guide; 9 – exit pipeline of in-fuel-rod coolant 1; 10 – entrance pipeline of in-fuel-rod coolant 1; 11 - exit pipeline of in-fuel-rod coolant 2; 12 - entrance pipeline of in-fuel-rod coolant 2; 13 - entrance pipeline of inter-hull coolant 3, 14 - exit pipeline of inter-hull coolant 3, 15 - entrance pipeline of hydraulic liquid; 16 - exit pipeline of hydraulic liquid; 17 – 24 – cylindrical reactor bearing construction elements)}
\label{fig01}
\end{figure*}

In order to reduce the radiation damage of the construction materials (fuel rod shell) significantly, the fuel movement speed relative to fixed fuel rod shell (along the channel-like active zone where the fuel rods shells are the most close to nuclear fuel channel shells (Fig.~\ref{fig01})) must satisfy the following relation:

\begin{equation}
v_{az} \geqslant \frac{l_{az}}{t_{fuel~burn}} = \frac{l_{az} v_{nucl. burn}}{l_{fuel}}
\label{eq05}
\end{equation}

\noindent where $l_{az}$ is the active zone length (fixed constructive part of the active zone in Fig.~\ref{fig01}).

The TWR construction materials campaign time may be written in the following form:

\begin{equation}
t_{TWR} ^{camp metal} \approx \frac{l_{burning~wave}}{v_{az} + v_{nucl. burn.}} 
\sim \frac{\lambda _{diff.~neutrons}}{v_{az} + v_{nucl.burn}}
\label{eq06}
\end{equation}

\noindent where $l_{burning wave}$ is the burning wave width (local burning zone), $\lambda _{diff. ~neutrons}$ is the neutron diffusion length.

The expression~(\ref{eq06}) may be transformed using~(\ref{eq05}) into:

\begin{equation}
t_{TWR} ^{camp.metal.} \sim \frac{l_{burning~wave}}{v_{az} + v_{nucl. burn.}} 
\leqslant \frac{l_{burning~wave}}{(l_{az} / l_{fuel} +1) v_{nucl. burn}}
\label{eq07}
\end{equation}

From~(\ref{eq07}) considering the estimate of the TWR metals campaign~(\ref{eq03}) we obtain:

\begin{align}
\frac{l_{fuel}}{l_{az}} & \geqslant \frac{v_{nucl. burn}}{l_{burning~wave}} (t_{TWR} ^{camp.metal.} +1) \sim \nonumber \\
& \sim \frac{2.31 \cdot 10^{-5} ~cm/s}{100 ~cm} \cdot 4.5 \cdot 10^5 ~s  \sim 10^{-1} 
\label{eq08}
\end{align}

\noindent where $v_{nucl. burn} = 2.31 \cdot 10^{-5} ~cm/s$, $l_{burning~wave} \sim 100 ~cm$.

Therefore, according to~(\ref{eq05}) and~(\ref{eq07}), the following nuclear fuel movement speed may be estimated as:

\begin{equation}
v_{az} \geqslant \frac{l_{az}}{l_{fuel}} v _{nucl. burn.} = 10 \cdot v_{nucl. burn.}
\label{eq09}
\end{equation}

Consequently, according to~(\ref{eq07}), for $l_{fuel} \sim 5~m$ we obtain $l_{az} \sim 50~m$, and according to (5), for $v_{nucl.burn.} =  2.31 \cdot 10^{-5} ~cm/s$ we obtain $v_{az} \sim 10 \cdot 2.31 \cdot 10^{-5} ~cm/s \sim 2.31 \cdot 10^{-4} ~cm/s$, and the TWR campaign time

\begin{align}
& t_{TWR} ^{camp. metal.} \sim \frac{l_{az}}{v_{az}} \sim \nonumber \\
& \sim \frac{5 \cdot 10^3 ~cm}{ 2.31 \cdot 10^{-4} ~cm/s \cdot 3\cdot 10^7 ~s/year}
\sim 0.72 ~years
\end{align}

From this example calculation it is clear that the main physical parameters determining the spatial and temporal parameters of the possible TWR construction are $\Phi_{TWR}$, $v_{nucl. burn.}$ and $l_{burning wave}$.

It should be noted that these parameters may be calculated by mathematical modeling of the wave nuclear burning kinetics.

Obviously, since the fuel movement speed along the active zone must satisfy expression~(\ref{eq05}), it may be provided by a technical implementation during the reactor construction. It may even be increased in case the greater reduction of radiation impact at the fuel rod shell is required.

For the obtained estimates the following neutron flux density was also obtained:

\begin{equation}
\Phi_{TWR} \sim 10 ^{17} ~neutrons / cm^2 \cdot s
\end{equation}

This value was chosen by modeling results:
\begin{itemize}
\item According to our results for fast U-Pu cylindrical reactor (diameter 70~cm, length 400~cm) $\Phi \sim 10^{19}~neutrons / cm^2 \cdot s$~\citep{ref21,ref22,ref23};
\item According to Fomin's group data for Th-U reactor $\Phi \sim 10^{16} ~neutrons / cm^2 \cdot s$~\citep{ref24,ref25,ref26}.
\end{itemize}

Considering that flux density will significantly reduce under "softening" of the spectrum, and wave reactor-transmutator for nuclides accumulated due to reactor fuel burning that create the highest hazard for biosystem, ideally should operate on neutrons with energy around 1~keV (intermediate neutrons), the abovementioned estimate for the neutron flux was chosen.

The burning wave width (local burning zone) $l_{burning~wave}$ was chosen 100~cm for the estimate~\citep{ref21,ref22,ref23,ref24,ref25,ref26}. 

Let us note that neutron diffusion length will decrease in case of the spectrum softening due to increase of the nuclear reactions cross-sections. And so does the burning zone length $l_{burning~wave}$. Thus according to~(\ref{eq07}) and~(\ref{eq08}), the nuclear fuel movement speed $v_{nucl.burn.}$ and the active zone length $l_{az}$ may be reduced.

Let us also note that the estimates were made for the hardest conditions of the materials operation: high burnout and minimal burning speed (non-enriched technical and natural uranium).

Hence in case the neutron-nuclear burning kinetics is implemented in such a way that the flux density is high and the burning wave speed is small, i.e. integral fluence at materials exceeds the currently acceptable level, then this problem may be solved by increasing the fuel local burning zone movement speed relative to edges/channels of the fuel rods shells.

To achieve this, according to the above said, we must ensure that the fuel movement speed along the fuel channel $v{az}$ is greater than the fuel burning wave speed $v_{nucl. burn.}$, according to the estimate obtained above by technical implementation. As shown above, these estimates for the chosen parameters are: $v_{az}  \sim 2.31 \cdot 10^{-4} ~cm/s$ and  $v_{nucl.burn.} \sim 2.31 \cdot 10^{-5} ~cm/s$. I.e. if one is able to technically implement the fuel movement speed along the channel just ten times greater than the neutron-nuclear fuel burning wave speed, then one would solve the problem of radiation stability of the fuel rod shells.

For comparison let us give the estimates for the case when an absorbing moderator layer is located between the fissile fuel and the fuel channel shell metal. It reduces neutron flux density at fuel channel shell e.g. two times relative to the flux density in previous calculations, i.e. $\Phi _{TWR} \sim 5 \cdot 10^{16} ~ neutrons / cm^2 \cdot s$. 

Then, according to~(\ref{eq03}), for the TWR metal campaign time we obtain:

\begin{align}
t_{TWR} ^{camp.metal.} & \sim \frac{500 \cdot 10^{14} \cdot 3}{100 \cdot 5 \cdot 10^{16}} years \sim \nonumber \\
& \sim 3.0 \cdot 10^{-2} ~years \sim 9.0 \cdot 10^5 ~s
\label{eq10}
\end{align}

And correspondingly according to~(\ref{eq08}) and~(\ref{eq09}):

\begin{equation}
\frac{l_{fuel}}{l_{az}} \geqslant \frac{2.31 \cdot 10^{-5} ~cm/s}{100 ~cm} \cdot 9.0 \cdot 10^5 ~s \sim 0.22
\label{eq11}
\end{equation}

and

\begin{equation}
v_{az} \geqslant \frac{l_{az}}{l_{fuel}} v_{nucl.burn.} \approx 5 \cdot v_{nucl.burn.}
\label{eq12}
\end{equation}

Therefore, for the considered variant according to~(\ref{eq11}) in case $l_{fuel} = 5 ~m$ we obtain $l_{az} = 25 ~m$ and according to~(\ref{eq12}) in case $v_{nucl.burn.} \sim 2.31 \cdot 10^{-5} ~cm/s$ cm/s we obtain $v_{az} = 5 \cdot 2.31 \cdot 10^{-5} ~cm/s \sim 1.16 \cdot 10^{-4} ~cm/s$, and the wave reactor campaign time

\begin{align}
& t_{TWR} ^{camp.metal.} \sim \frac{l_{az}}{v_{az}} \sim \nonumber \\
& \sim \frac{2.5 \cdot 10^3 ~cm}{1.16 \cdot 10^{-4} ~cm/s \cdot 3\cdot 10^7 ~s/year} \sim 0.72 ~years
\end{align}

Therefore we reduce the active zone length two times from 50~m in the first variant to 25~m for the second one. This is important because it significantly reduces the possible reactor size and increases the practical implementability of such reactor.

The Fig.~\ref{fig01} presents a concept scheme of the channel-type reactor with one burning fuel rod. Here the reactor hull shape is cylindrical. This is a case of homogeneous active zone of large diameter ($\sim$1-3~m). In the given scheme the fuel movement with the given speed is performed by the movement of the bearing frame implemented as a moving part of a hydraulic system. Such construction fits well the reactor prototype enabling experimental testing of burning wave kinetics and all the principal physical and construction parameters. Let us note that there are three coolants in the proposed construction -- two in-fuel-rod coolants (coolant 1 and 2 in the scheme), while different coolants may be used; and one inter-hull coolant (coolant 3).

The Fig.~\ref{fig01} does not present the aspect ratio for the bearing frame of the hydraulic system for the fuel movement precisely. Indeed, this figure rather corresponds to the case of a hydraulic system bearing frame implementation based on coaxial cylinders principle (radio antenna principle), which is widely used in hydraulic cranes and enabling the reduction of reactor construction size.

Of course, other known and appropriate engineer solutions may be used for the fuel movement system construction.

According to the estimates given above, it seems reasonable to choose the following geometrical parameters for the long-campaign operating reactor prototype project: $l_{fuel} \sim 5~m$, $l_{az} \sim 25 \div 50 ~m$, hull diameter $\sim 20 \div 30~m$, height $\sim 30 \div 60 ~m$.

The fuel used may be:
\begin{itemize}
\item Metallic $^{238}$U with small Molybdenum addition (up to 10\%) for stabilization of cubic uranium lattice up to room temperatures (melting temperature 1406~K);
\item Metallic $^{238}$U with small Chromium addition (up to 8\%) for stabilization of the radiation form change of the alloy (melting temperature $\sim$1400~K);
\item Natural uranium as uranium-based metallic alloys;
\item $^{238}$U dioxide (melting temperature 2820~K);
\item MOX-fuel;
\item Cermet fuel;
\item Uranium carbides;
\item Nitride fuel;
\item Spent nuclear fuel of many nuclear reactor types, e.g. dioxide, nitride fuel, dispersive type fuel etc.;
\item Nuclear fuel based on $^{232}$Th and $^{238}$U.
\end{itemize}

The fuel problem apparently requires additional research.

Gases used in gas reactors, water and metallic coolants used in reactors (e.g. lithium, natrium, stanum, lead, bismuth, mercury, lead-bismuth mixture, and their complexes) may be used as coolants in different coolant pump designs (Fig.~\ref{fig01}).

The specific type of coolant or coolant system providing for thermal-and-physical reactor parameters given by the wave reactor construction requirements specification may be determined only by proper thermal-and-physical calculations and investigations that are yet only planned.

The reactor steel HT9, X18H10T and others may be used as fast reactor construction materials.

\begin{figure*}[htbp]
\begin{center}
\includegraphics[width=15cm]{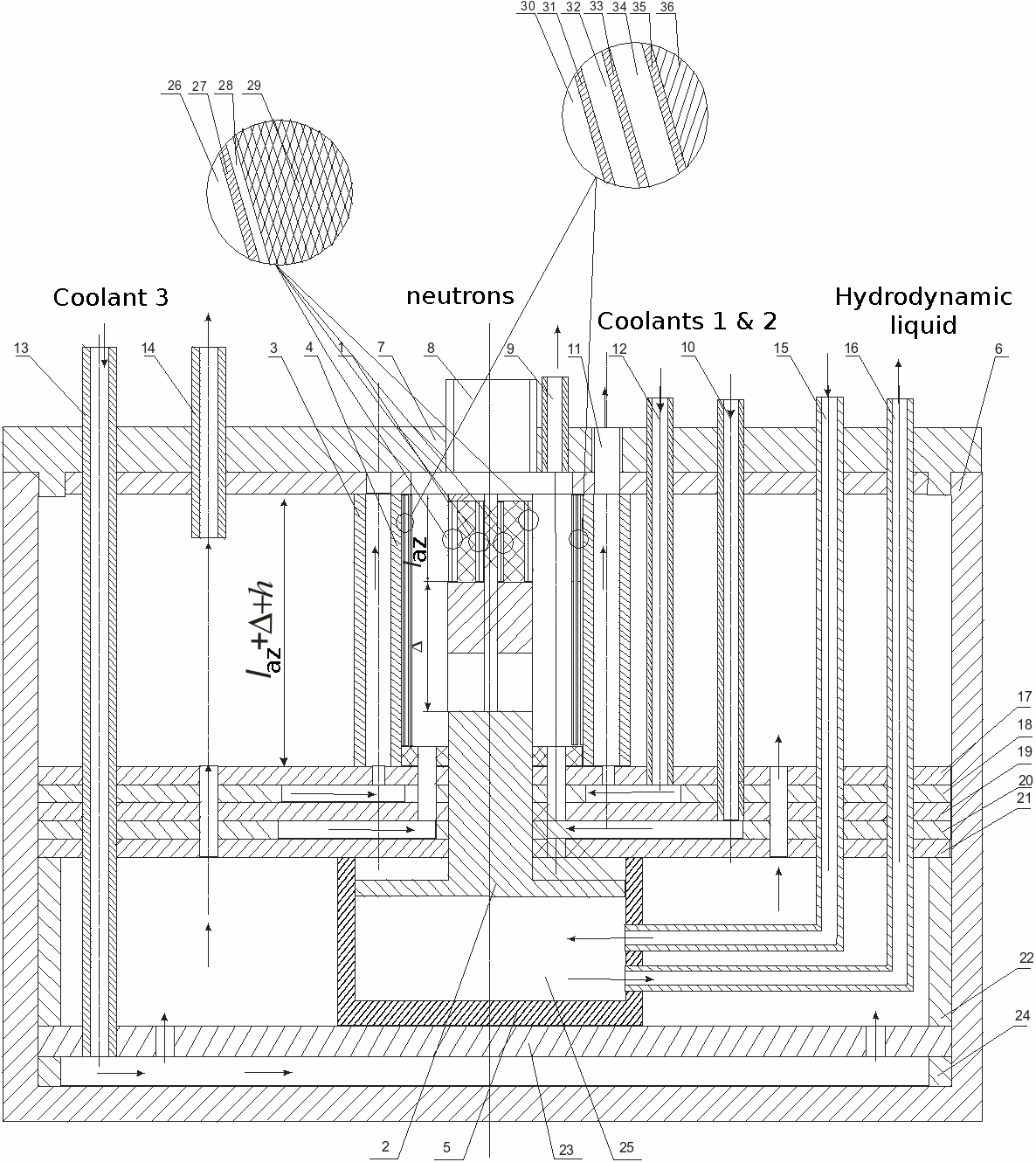}
\end{center}
\caption{Concept scheme of a TWR with one burning fuel rod. Homogeneous large-diameter ($\sim$1-3~m) active zone variant.
(1 – the nuclear fuel; 2 – moving construction of the hydraulic fuel movement system; 3 – metal of the fuel channel 2; 4 – metal of the fuel channel 1 shell; 5 – hydraulic liquid reservoir hull; 6 – reactor hull; 7 – reactor hull cover; 8 – neutron guide; 9 – exit pipeline of in-fuel-rod coolant 1; 10 – entrance pipeline of in-fuel-rod coolant 1; 11 - exit pipeline of inter-channel coolant 2; 12 - entrance pipeline of inter-channel coolant 2; 13 - entrance pipeline of inter-hull coolant 3, 14 - exit pipeline of inter-hull coolant 3, 15 - entrance pipeline of hydraulic liquid; 16 - exit pipeline of hydraulic liquid; 17 – 24 – cylindrical bearing, constructive elements of the reactor; 25 – hydraulic liquid; 26 – coolant 1; 27 – protection shell metal; 28 – beryllium moderator; 29 – fuel; 30 – coolant 1; 31 - protection shell metal; 32 - beryllium moderator; 33 - protection shell metal; 34 – carbon; 35 - protection shell metal; 36 - fuel channel 1 shell metal)}
\label{fig02}
\end{figure*}

Let us note that the coolant pipeline system may be used in the prototype to implement neutron reflector, e.g. water or beryllium, and also for accommodation of moderator-absorbent layer near the fuel rod edge to reduce the fuel rod shell radiation damage (Fig.~\ref{fig02}). This may enable reduction of fuel movement speed and, respectively, the length of the active zone $l_{az}$.

Indeed, the radiation damage of the fuel rod shell construction materials may be reduced by reduction of the neutron flux achieved by placing the specific quantity of a specially chosen substance with proper characteristics of neutron moderator and absorbent between the fissile material and fuel rod shell. As a result of the neutrons moderation during the interaction with nuclei of the moderator the neutron speed reduces leading to neutron flux density reduction. Due to neutron capture by absorbent the neutron concentration reduces, which also leads to neutron flux density reduction. In nuclear reactors physics the moderation efficiency and moderation coefficient are used as the quantitative moderator characteristics.

The moderator substance must posses a high moderating efficiency and low moderation coefficient for the optimal solution of the problem of neutron flux density reduction by passing through moderator substance. The efficiency of the neutron flux density reduction also depends on moderator-absorbent nuclei concentration i.e. on moderator-absorbent substance density which may be changed by its thermodynamical parameters such as volume, pressure and temperature.

Assuming that nuclear fuel, shell construction material and moderator-absorbent in-between have cylindrical shape, the estimate of radial width of moderator-absorbent layer required for the given neutron flux density reduction may be calculated as follows.

Suppose the flux density under moderation reduces due to neutron energy reduction from $E_{fuel}$ (energy of neutrons released from the nuclear fuel) to $E_{shell}$ (energy of neutrons at the shell after the moderator). Let atomic number of moderator-absorbent substance to be $A$. The logarithmic mean neutron energy loss (attenuation) during its moderation may be calculated as~\citep{ref69,ref70}:

\begin{equation}
\xi = 1 + \frac{(A - 1)^2}{2A} \ln {\frac{A - 1}{A + 1}}
\label{eq13}
\end{equation}

The average impact ratio of the neutron being moderated by the moderator nuclei, required for neutron energy reduction from $E_{fuel}$ to $E_{shell}$ equals~\citep{ref69,ref70}:

\begin{equation}
n = \frac{1}{\xi} \ln {\frac{E_{fuel}}{E_{shell}}}
\label{eq14}
\end{equation}

Neutron free path in the moderator is~\citep{ref69,ref70}:

\begin{equation}
\lambda = \frac{1}{\Sigma_s + \Sigma_a}
\label{eq15}
\end{equation}

\noindent where $\Sigma_s$ is the neutron scattering macroscopic cross-section and $\Sigma_a$ is the neutron absorption macroscopic cross-section.

The estimates of scattering and absorption cross-sections may be obtained by expressions~\citep{ref69,ref70}:

\begin{equation}
\Sigma_s \approx \hat{\sigma_s} N_{moderator} ~~~ and \Sigma_a \approx \hat{\sigma_a} N_{moderator}
\label{eq16}
\end{equation}

\noindent where $\sigma_s$ and $\sigma_a$ are the neutron scattering and absorption microscopic cross-sections respectively, averaged over the energy interval from $E_{fuel}$ to $E_{shell}$, $N_{moderator} = \rho N_A / \tilde{A}$ is the moderator nuclei density, where $\rho$ is the moderator density, $N_A$ is the Avogadro number, $\tilde{A}$ is the molar mass of the moderator.

Then moderator-absorbent layer width  required for the given neutron flux density reduction may be estimated as follows:

\begin{equation}
R_{moderator} \approx \lambda n
\label{eq17}
\end{equation}

Below in Table~\ref{tab2} we present the characteristics of some known moderators and the corresponding moderator layer width estimates made for neutrons moderating from $E_{fuel}$ = 1.0~MeV to $E_{shell}$ = 0.1~MeV. ENDF-VII data on cross-sections were used for calculations.

\begin{figure*}[tbp]
\begin{center}
\begin{minipage}{0.49\linewidth}
\centering
\includegraphics[width=6cm]{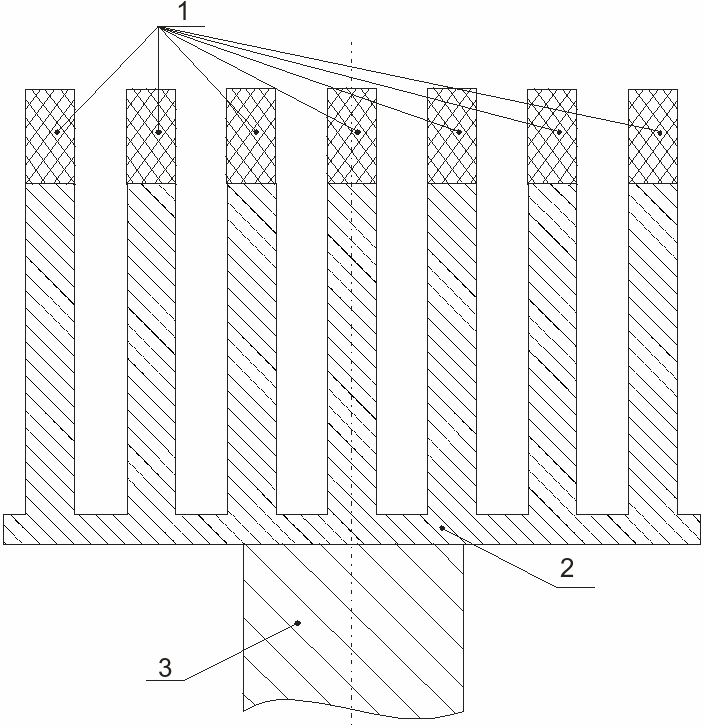}
\end{minipage}
\hfill
\begin{minipage}{0.49\linewidth}
\centering
\includegraphics[width=8cm]{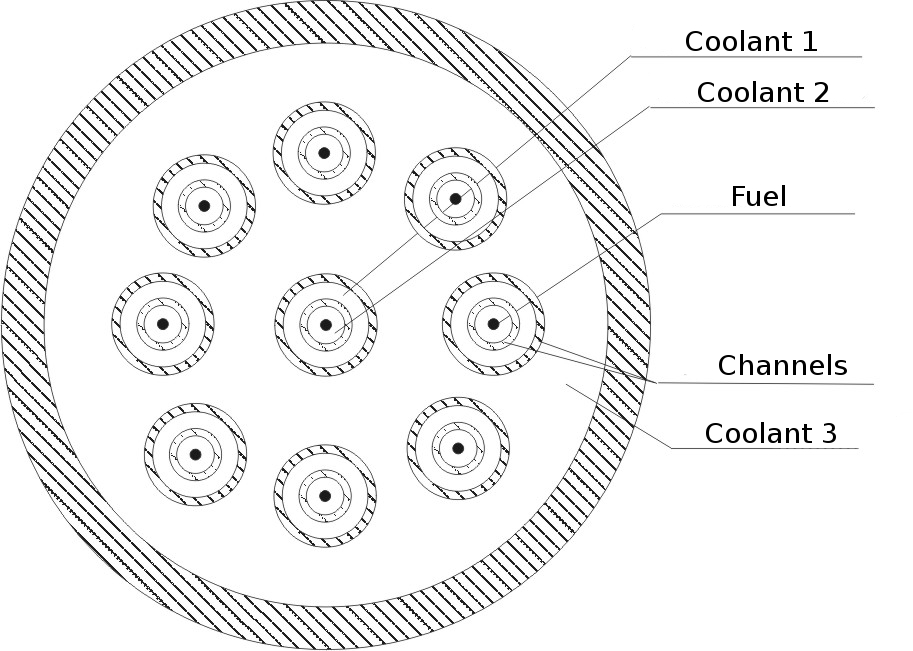}
\end{minipage}
\end{center}
\caption{\textbf{Left panel:} A scheme explaining principles of possible fuel rods placement over moving part of the fuel rods hydraulic movement system (1 - fuel; 2 and 3 - hydraulic fuel movement system bearing frame).\newline
\textbf{Right panel:} Reactor active zone section scheme (heterogeneous active zone case).}
\label{fig03}
\end{figure*}

The scheme presented at Fig.~\ref{fig01} may be easily generalized for a bigger number of fuel rods. For example, Fig.~\ref{fig03} presents a possible scheme of seven fuel rods placement on bearing frame driven by hydraulic system of the fuel movement. The fixed fuel rods shells and coolant construction is also easily generalized for bigger number of fuel rods by simply replicating its construction presented in Fig.~\ref{fig01} or Fig.~\ref{fig02}. In such a way one obtains a heterogeneous active zone.

Several possibilities for the reactor implementation arise at this point.

First, an implementation simply generalizing homogeneous large-diameter active zone implementation is possible. I.e. a set of several such active zones burning independently from each other, or even sequentially one after another given the corresponding design. In case of simultaneous implementation of independent homogeneous active zones we obviously obtain high-power reactor with relatively short campaign time in comparison to implementation with sequential burning of such active zones. Such a reactor would be promising for burning of spent nuclear fuel in large amounts.

\begin{table*}
  \begin{footnotesize}
  \begin{center}
  \begin{tabular}{|c|c|c|c|c|c|c|c|c|}
    \hline
     \multirow{4}{1.5cm}{Moderator} &  \multirow{4}{1cm}{Mass number, $A$} & \multirow{4}{1.75cm}{Mean logarithmic energy, $\xi$} &  \multirow{4}{1cm}{density, $\rho$, g/cm$^3$} & \multirow{4}{1.9cm}{Impacts required for moderating, $n$} &  \multirow{4}{1.5cm}{Neutron mean free path, $\lambda$} & \multirow{4}{1.75cm}{Moderating ability, $\xi \Sigma _s$, cm$^{-1}$} & \multirow{4}{1.75cm}{Moderating coefficient,  $\xi \Sigma_s / \Sigma _a$}& \multirow{4}{1.5cm}{Moderator layer width, $R_{mod}$, cm} \\
    & & & & & & & &\\
    & & & & & & & &\\
    & & & & & & & &\\
    \hline    
    Be &   9 & 0.21 & 1.85 & 11 & 1.39 & 0.151 & 15100 & 15.3 \\
    \hline
    BeO &   25 &  & 2.96 &  &  & & &  \\
    \hline
    C   & 12 & 0.158 & 1.60 & 15 & 3.56 & 0.044 & & 53.4 \\
    \hline
    H$_2$O & 18 & 0.924 & 1.0 & 2.5 & 16.7 & 0.055& & 41.6 \\
    \hline
    H$_2$O + 10\%B& & & & 2.5 &10.0 & &  & 25.0 \\
    \hline
    D$_2$O & 20 & & 1.1 & & & & &  \\
    \hline
    H & 1 & 1.0 & $0.09\cdot 10^{-4}$ & 2.3 & $0.37 \cdot 10^{4}$&  $2.65 \cdot 10^{-4}$& $0.62 \cdot 10^{5}$& $0.85 \cdot 10^4$  \\
    \hline   
    He & 4 & 0.425 & 0.00018 & 5.41 & $10^4$ & & & $5.41 \cdot 10^4$ \\
    \hline
    He & 4 & 0.425 & 0.18 & 5.41 & 11.2 & & & 60.7 \\
    \hline
    CO$_2$ &   &   & $1.98 \cdot 10^{-3}$ &   &   & & &   \\
    \hline
    Pb & 207  & 0.01  & 11.3 & 230  & 4.27  &0.0023 & $0.75 \cdot 10^3$ &   982.1\\
    \hline
    Hg &  200 & 0.01  & 13.6 &  230 &  3.01 & 0.0033& 15.71&  692.8  \\
    \hline
    Bi &  209 &  0.01 & 9.8 &  230 &  5.08 & 0.0020& 16.02& 1167.5  \\
    \hline
    Sn & 118  &  0.02 & 7.4 &  115 & 4.2  & 0.0048& 3.2&  483.2 \\
    \hline
  \end{tabular}
  \end{center}
  \end{footnotesize}
\caption{Moderating and absorbing properties of some substances, moderator 
         layer width estimate for moderating neutron from $E_{fuel} = 1.0 ~MeV$ 
	 to $E_{mod} = 0.1 ~MeV$.}
\label{tab2}
\end{table*}

In case of sequential burning design one obtains a source of relatively lower power than in the previous case and with significantly longer campaign time.

Second, the implementation of the reactor in a form of the thin fuel rods set in a collective burning zone is possible. This is analogous to traditional structure of the operating heterogeneous reactors.

Let us note that the fixed reactor active zone part becomes very similar to channel type reactors in this case, in particular to LWGR reactors (see Fig.~\ref{fig03}). This also enables one to benefit from the advantages of channel reactors such as increased thermal and physical parameters (temperature and pressure) inside the channels.

Let us also note that the coolant entrance channel system presented in Fig.~\ref{fig01} and Fig.~\ref{fig02} may be replicated and spread by height (length) of the fixed active zone part. This would solve the problem of the increased hydraulic resistance for a long active zone. Indeed, in such approach one can increase the quantity of coolant entrance channels reducing distance between them by height (length) of the active zone to provide for technically implementable coolant hydraulic resistance in the pumping channels.

The examples of the TWR active zone designs given here are clearly only hypothetically possible. Later we need to investigate the kinetics of these reactors by mathematical modeling. 

Based on the obtained model results we prepare the requirements specifications, the assignment for developing of the prototypes and development prototypes, design them, create the development prototypes and test them. Only after all that the operating reactors may be developed.

According to the above given estimates it seems that acceptable geometrical parameters of the operating reactor with long campaign may be chosen as: $l_{fuel} \sim 5~m$, $l_{az} \sim 25 \div 50 ~m$, hull diameter $\sim 30~m$, height $\sim 30 \div 60 ~m$.

\begin{figure*}[htbp]
\begin{center}
\begin{minipage}{0.49\linewidth}
\centering
\includegraphics[width=6cm]{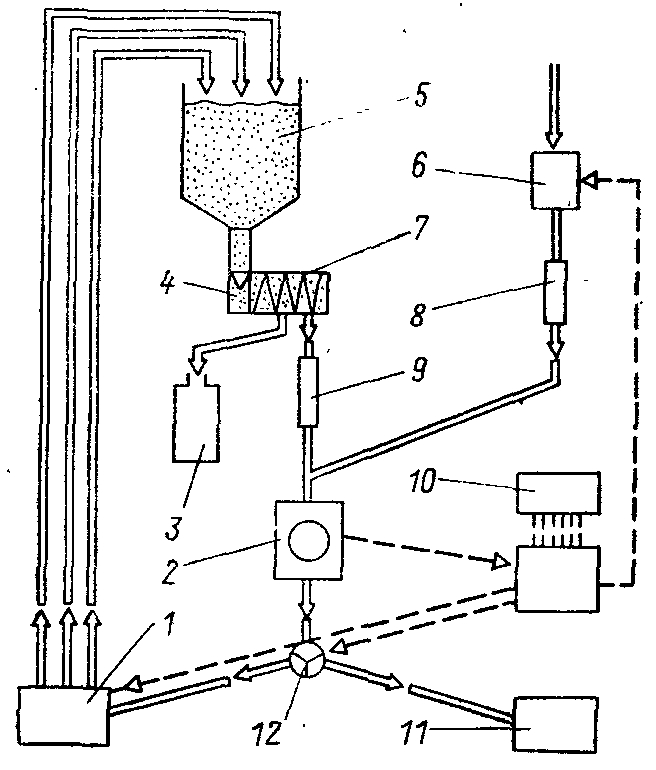}
\end{minipage}
\hfill
\begin{minipage}{0.49\linewidth}
\centering
\includegraphics[width=6cm]{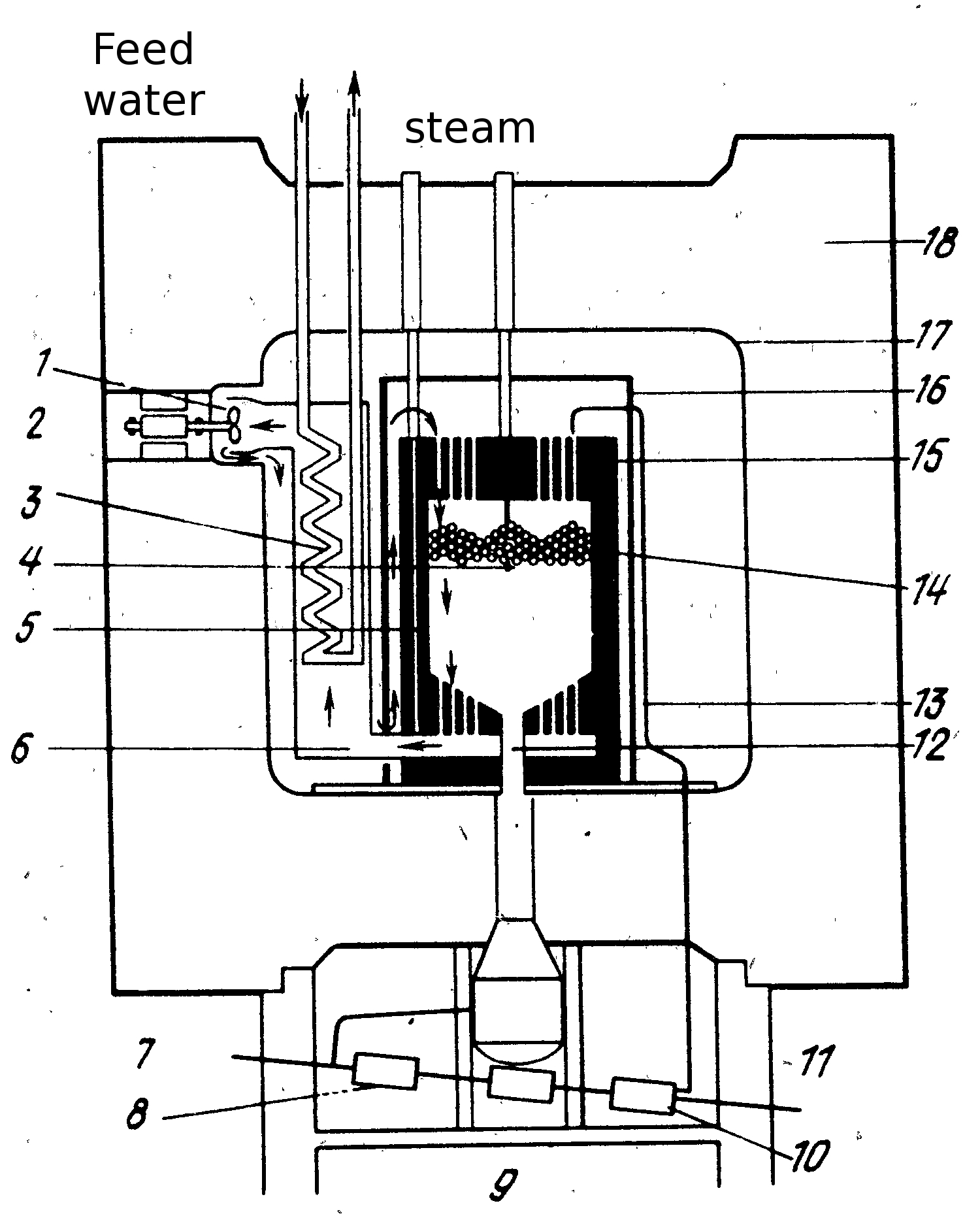}
\end{minipage}
\end{center}
\caption{High-temperature reactor with spherical fuel elements construction scheme~\citep{ref71,ref72,ref73}.
\textbf{Left panel:} high temperature gas cooled reactor scheme with fill-in active zone: 1 - fuel elements elevator; 2 - fuel elements investigation system and burn-out level measurement; 3 - damaged elements rods vault; 4 - accumulator; 5 - active zone; 6 - fuel elements feed system; 7 - stepwise fuel elements separator; 8,9 - intermediate screw conveyors; 10 - process control computer; 11 - fuel elements withdrawal system; 12 - directing device.\newline
\textbf{Right panel:} HRB reactor scheme with spherical fuel elements: 1 – gas blower; 2 – gas  blower gear; 3 – steam generator; 4 – absorbing reflector rod; 7 – spherical fuel elements feed line; 8 – accumulating block; 9 – fuel burn-out level measuring device; 10 – spheres sorting and transport device; 11 – spheres unloading device; 12 – pipe for spheres unloading; 13 – fuel feed channel; 14 – spheres fill-in; 15 – reflector; 16 – heat shield; 17 – coating; 18 – hull made of preliminarily strained armored concrete.}
\label{fig05}
\end{figure*}

It seems very promising to use a set of $^{238}$U spheres filled into a fixed cylinder of the fuel rod shell instead of the metallic $^{238}$U rod (Fig.~\ref{fig01}, Fig.~\ref{fig03}) fixed on moving bearing frame implemented as moving part of the hydraulic system. Of course, the micro fuel rods or spherical fuel elements analogous to spherical fuel elements of high-temperature gas reactors (e.g. analogous to THTR-300~\citep{ref71,ref72} and VTGR-500 (high temperature gas cooled reactor)~\citep{ref73}) (Fig.~\ref{fig05}) may be used as a fuel. In the considered case they slowly move along the fuel rod shell following the movement of the bearing framework platform, implemented in form of a moving part of the hydraulic system.

Naturally, we speak of $^{238}$U spheres with sizes appropriate for the required constructive allowances for fuel rod shell diameter and moving bearing framework platform diameter to prevent them falling outside the platform. These may be $^{238}$U spheres with protective coating made of Si, typical for micro-fuel rods, and probably with carbon layer for the neutron spectrum softening. Let us note that these spheres may be of different diameter: e.g. a layer of larger diameter spheres may be located below to meet construction demands, and the micro-fuel elements spheres above them to increase the fission environment density. However, all these problems have very little principal significance at the moment and are to be solved during the specific construction development. It is important for the process of fuel spheres moving down not to interfere with wave neutron-nuclear fuel burning process.

Let us also note that the concept of burning fuel movement relative to fuel rods shell made of constructive materials considered in this patent also corresponds to a possible active zone construction based on the principles already implemented in known high-temperature gas power reactors (reactors THTR-300~\citep{ref71,ref72} and VTGR-500 (high temperature gas cooled reactor)~\citep{ref73} (Fig.~\ref{fig05})). Their active zone is a hull with cone-like bottom with a hole in the center. The spherical fuel elements are filled in from above and while burning-out, they fall through a hole from the active zone to a spent fuel elements container. Such construction of the TWR may require burn-up of the fuel both in the upper part of the active zone and in the lower part, which requires locating the neutron source for the nuclear burning wave burn-up inside the reactor hull or below it.

\begin{figure*}[htbp]
\begin{center}
\includegraphics[width=8cm]{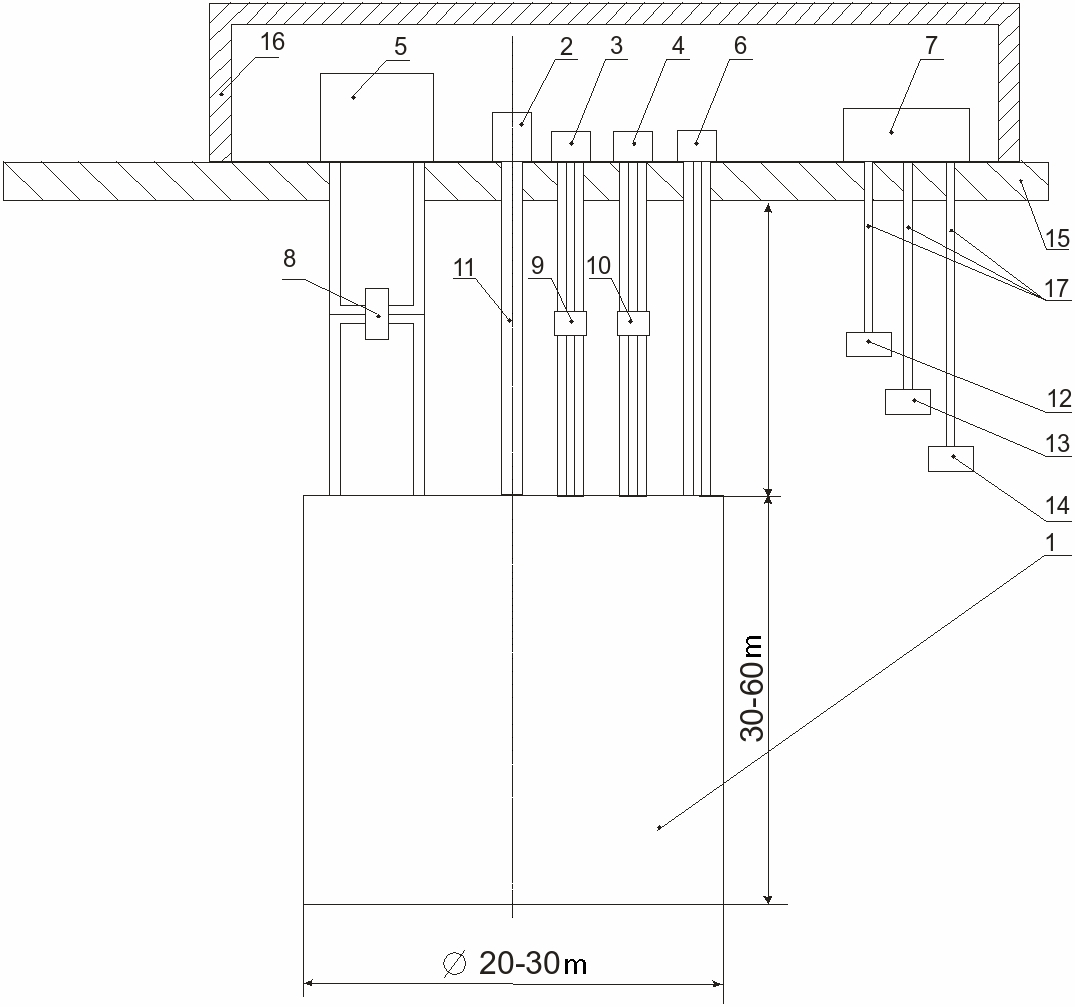}
\end{center}
\caption{Wave reactor and reactor equipment arrangement scheme.
(1 - reactor; 2 - accelerator or impulse reactor; 3 - coolant 1 pump system and heat takeoff; 4 - coolant 2 pump system and heat takeoff; 5 -  coolant 3 pump system and heat takeoff; 6 - hydraulic liquid pump system; 7 - neutrino control system; 8 - coolant 3 heat exchanger; 9 - coolant 1 heat exchanger; 10 - coolant 2  heat exchanger; 11 - neutron guide; 12,13,14 - neutrino detectors; 15 - ground surface; 16 - reactor hull; 17 - neutrino control system communication line).}
\label{fig06}
\end{figure*}

The Fig.~\ref{fig06} presents the reactor and reactor equipment location scheme. It gives the size of the reactor prototype. A particle accelerator (e.g.~\citep{ref74}) or impulse nuclear reactor (e.g.~\citep{ref75,ref76}) are proposed (see Fig.~\ref{fig06}) as an external neutron source. For the sake of the neutron-nuclear burning wave initiation (burn-up) optimization the upper fuel part may be enriched by some fissionable nuclide to such amount that this enriched area would be in under-critical state. Therefore the proposed channel type TWR construction is a reactor with internal safety~\citep{ref17,ref17a}.

It should be noted that both the TWR project and the spent nuclear fuel processing reactor are apparently single-load reactor projects. After burning the fuel the reactor installation is buried. An implementation of the remote neutrino control of the neutron-nuclear burning wave kinetics~\citep{ref23} (Fig.~\ref{fig06}) is obviously required.

\section*{Conclusions}

The basic design of the fast uranium-plutonium nuclear TWR with a softened neutron spectrum is developed. It solves the problem of the fuel claddings material radiation resistance. This reactor may also work as a processor of the spent nuclear fuel.

\bibliographystyle{unsrtnat}
\bibliography{channelTWR}

\end{document}